\pdfoutput=1
\documentclass[sigplan,screen,nonacm]{acmart}

\usepackage{amsmath}  
\usepackage{algorithmic}
\usepackage{graphicx}
\usepackage{textcomp}
\usepackage{xspace}

\usepackage{subcaption}
\usepackage{enumitem}
\usepackage{acronym}
\usepackage{pifont}
\usepackage{algorithmic}
\usepackage{tikz}
\usepackage[utf8]{inputenc}
\usepackage[linesnumbered,ruled,vlined]{algorithm2e}
\usepackage[switch]{lineno}
\usepackage{xspace}
\usepackage{booktabs}
\usepackage{comment}

\usepackage{textcomp}
\usepackage{array}
\usepackage{wrapfig}

\newcommand{\cmark}{\ding{51}}%
\newcommand{\xmark}{\ding{55}}%
\newcommand{\na}{--}%

\usepackage{hyperref}
\hypersetup{
    colorlinks=true,  
    linkcolor=blue,   
    citecolor=red,    
    filecolor=cyan,   
    urlcolor=magenta  
}

\definecolor{ForestGreen}{RGB}{34,139,34}
\definecolor{amber}{rgb}{1.0, 0.49, 0.0}

\usepackage[labelfont=bf]{caption}
\usepackage[T1]{fontenc}
\usepackage{fancyvrb}

\SetKwSty{text}
\SetArgSty{texttt}
\SetDataSty{}
\SetKwInput{KwInput}{\textcolor{blue}{\textbf{Input}}}
\SetKwInput{KwOutput}{\textcolor{blue}{\textbf{Output}}}
\SetKwIF{If}{ElseIf}{Else}{\textcolor{blue}{\textbf{if}}}{\textcolor{blue}{\textbf{then}}}{\textcolor{blue}{\textbf{else if}}}{\textcolor{blue}{\textbf{else}}}{\textcolor{blue}{\textbf{endif}}}
\SetKwFor{For}{\textcolor{blue}{\textbf{for}}}{\textcolor{blue}{\textbf{do}}}{\textcolor{blue}
{\textbf{endfor}}}
\SetKwFunction{Length}{length}
\SetKwFunction{Max}{max}
\SetKwProg{Fn}{\textcolor{blue}{\textbf{function}}}{}{}
\SetKw{KwRet}{\textcolor{blue}{\textbf{return}}}
\SetKwComment{Comment}{/* }{ */}
 
\acrodef{ML}[ML]{Machine Learning}

\acrodef{NSF}[NSF]{National Science Foundation}

\acrodef{AI}[AI]{Artificial Intelligence}

\acrodef{FL}[FL]{Federated Learning}

\acrodef{CL}[CL]{Critical Learning}

\acrodef{AC}[AC]{Attacking-Critical}

\acrodef{CAGR}[CAGR]{compound annual growth rate}

\acrodef{CCT}[CCT]{Center for Computation and Technology}

\acrodef{SLO}[SLO]{service level objective}

\acrodefplural{SLOs}[SLO]{service level objectives}

\acrodef{RL}[RL]{reinforcement learning}

\acrodef{DRL}[DRL]{deep reinforcement learning}

\acrodef{VM}[VM]{virtual machine}

\acrodefplural{VM}[VMs]{virtual machines}

\acrodef{ITC}[ITC]{Innovation \& Technology Commercialization}

\acrodef{DAG}[DAG]{directed acyclic graph}

\acrodefplural{DAG}[DAGs]{directed acyclic graphs}

\acrodef{SFA}[SFA]{single point authentication}

\acrodef{HPC}[HPC]{high-performance computing}

\acrodef{SBIR}[SBIR]{Small Business Innovation Research}

\acrodef{IoT}[IoT]{Internet of Things}

\acrodef{DML}[DML]{distributed machine learning}

\acrodef{GNN}[GNN]{graph neural network}

\acrodefplural{GNN}[GNNs]{graph neural networks}

\acrodef{BSR}[BSR]{backdoor success rate}

\acrodef{BTA}[BTA]{backdoor task accuracy}

\acrodef{ATT}[ATT]{App Tracking Transparency}

\acrodef{DNN}[DNN]{deep neural network}

\acrodef{DNNs}[DNNs]{deep neural networks}

\acrodef{KL}[KL]{Kullback–Leibler}

\acrodef{IaaS}[IaaS]{Infrastructure-as-a-Service}

\acrodef{CaaS}[CaaS]{Container-as-a-Service}

\acrodef{TRPO}[TRPO]{Trust Region Policy Optimization}

\acrodef{CPO}[CPO]{Constrained Policy Optimization}

\acrodef{PPO}[PPO]{Proximal Policy Optimization}

\acrodef{TV}[TV]{Total Variation}

\acrodef{PAC}[PAC]{Probably Approximately
Correct}

\acrodef{ACI}[ACI]{Azure Container Instances}

\acrodef{NLP}[NLP]{Natural Language Processing}

\acrodef{GAE}[GAE]{Generalized Advantage Estimation}

\acrodef{IS}[IS]{Importance Sampling}

\acrodef{CV}[CV]{coefficient of variance}

\acrodefplural{CV}[CVs]{coefficient of variances}

\acrodef{LLM}[LLM]{Large Language Model}

\acrodef{GNS}[GNS]{gradient noise scale}

\acrodef{SOTA}[SOTA]{state-of-the-art}

\acrodef{SSP}[SSP]{Stale Synchronous Parallel}

\acrodef{CDF}[CDF]{cumulative distribution function}

\acrodef{PDF}[PDF]{probability density function}

\acrodef{RPC}[RPC]{remote procedure call}

\acrodef{MARL}[MARL]{multi-agent reinforcement learning}

\acrodef{SARL}[SARL]{single-agent reinforcement learning}

\acrodef{MDP}[MDP]{Markov Decision Process}

\acrodef{CTDE}[CTDE]{centralized training \& decentralized execution}

\acrodef{MAPD}[MAPD]{Multi-Agent Policy Distance}

\acrodef{IPPO}[IPPO]{Independent Proximal Policy Optimization}

\acrodef{MPE}[MPE]{Multi-Agent Particle Environment}

\acrodef{SMAC}[SMAC]{StarCraft Multi-Agent Challenge}

\acrodef{DDPG}[DDPG]{Deep Deterministic Policy Gradient}

\acrodef{DQN}[DQN]{Deep Q-Network}

\acrodef{MAPPO}[MAPPO]{Multi-Agent Proximal Policy Optimization}

\acrodef{MADDPG}[MADDPG]{Multi-Agent Deep Deterministic Policy Gradient}

\acrodef{IQL}[IQL]{Independent Q-Learning}

\acrodef{KDE}[KDE]{Kernel Density Estimation}

\acrodef{VDN}[VDN]{Value-Decomposition Networks}

\acrodef{SAC}[SAC]{Soft Actor-Critic}

\acrodef{QMIX}[QMIX]{Monotonic Value Function Factorisation}

\acrodef{IDDPG}[IDDPG]{Independent Deep Deterministic Policy Gradient}

\acrodef{RNN}[RNN]{recurrent neural network}

\acrodef{NN}[NN]{neural network}

\acrodef{PER}[PER]{Prioritized Experience Replay}

\acrodef{RLHF}[RLHF]{Reinforcement Learning from Human Feedback}

\acrodef{RLAIF}[RLAIF]{Reinforcement Learning from AI Feedback}

\acrodef{GRPO}[GRPO]{Group Relative Policy Optimization}

\acrodef{DPO}[DPO]{Direct Preference Optimization}

\acrodef{SFT}[SFT]{Supervised Fine-tuning}

\acrodef{FSDP}[FSDP]{Fully Sharded Data Parallel}

\acrodef{MOE}[MOE]{Mixture-of-Experts}

\acrodef{RM}[RM]{Reward Model}

\newcommand{\sys}{\textit{GearUp}\xspace}

\begin{document}

\title{\sys: DAG-aware Acceleration for Serverless Inference Serving via Slack-guided Pre-loading}
\thanks{This work was performed when Qi Qi was an intern student advised by Dr.~Hao Wang at the IntelliSys Lab of Stevens Institute of Technology.}


\author{Qi Qi}
\affiliation{%
  \institution{University of Massachusetts Amherst}
  \city{Amherst}
  \state{Massachusetts}
  \country{USA}
}
\email{qqi@umass.edu}

\author{Yifan Sui}
\affiliation{%
  \institution{Hong Kong University of Science and Technology}
  \city{Hong Kong}
  \country{China}
}
\email{suiyifan.sjtu@gmail.com}

\author{Hanfei Yu}
\affiliation{%
  \institution{University of California, Berkeley}
  \city{Berkeley}
  \state{California}
  \country{USA}
}
\email{hanfeiyu@berkeley.edu}

\author{Hong Zhang}
\affiliation{%
  \institution{University of Waterloo}
  \city{Waterloo}
  \state{Ontario}
  \country{Canada}
}
\email{hongzhangblaze@gmail.com}

\author{Hao Wang}
\affiliation{%
  \institution{Stevens Institute of Technology}
  \city{Hoboken}
  \state{New Jersey}
  \country{USA}
}
\email{hwang9@stevens.edu}

\renewcommand{\shortauthors}{Qi Qi et al.}
%
% ...
%
\begin{abstract}

Serverless platforms increasingly host machine learning (ML) inference workflows, modeled as directed acyclic graphs (DAGs) of interdependent functions.
Beyond container startup, each ML node must load libraries, model weights, and CUDA kernels before it can run.
These costs accumulate along dependent DAG paths and place initialization on the end-to-end critical path.
Existing systems optimize one half of this problem.
DAG-agnostic pre-loading prepares nodes the request may never reach, wasting host and GPU memory.
DAG-aware serving routes and scales whole functions, leaving the loading stages inside a node untouched.

This paper presents \sys, a serverless DAG inference serving system that removes ML-node loading from the end-to-end critical path at low resource cost.
\sys profiles the loading stages of each node and schedules them inside the slack time the DAG exposes, hiding initialization behind upstream execution.
It reuses initialized instances within and across requests to avoid redundant cold starts.
A probabilistic, traffic-adaptive keep-alive policy sets how long idle instances are retained, balancing memory cost against cold-start risk under bursty load.
On the AMBER Alert and FINRA-ML workflows with NVIDIA L40S GPUs, \sys reduces median end-to-end latency by 64.5\% and 75.9\% against SMIless, for median speedups of $3.1\times$ and $10.0\times$ over cold execution.
Integrated GPU and host memory allocation fall by up to 83.9\% and 94.6\% against eager pre-loading in InstaInfer.
\sys achieves $4.2\times$ strong scaling on four workers.
\end{abstract}

\maketitle


\section{Introduction} \label{sec:intro}
Serverless computing offers fine-grained scaling and pay-as-you-go billing, and is increasingly used for Machine Learning (ML) inference, where workflows are modeled as Directed Acyclic Graphs (DAGs) of interdependent functions.
Platforms such as Kubeflow~\cite{kubeflow}, MLRun~\cite{mlrun}, and Airflow~\cite{apache_airflow} express inference pipelines in this form.
The same structure underlies video analytics~\cite{romero2021llama,poms2018scanner}, scientific workflows~\cite{roy2022daydream}, market surveillance~\cite{finra}, and agentic applications~\cite{langgraph,openai_agents_sdk}.
While attractive, serverless DAG-based inference faces a critical bottleneck from initialization overhead.
Beyond container startup, each ML node often requires expensive initialization steps, including loading libraries, model weights, and CUDA kernels.
Since these costs occur across dependent DAG stages, they accumulate along execution paths and significantly increase end-to-end latency, directly degrading user experience.

\begin{figure}[t]
  \centering
  \includegraphics[width=0.85\linewidth]{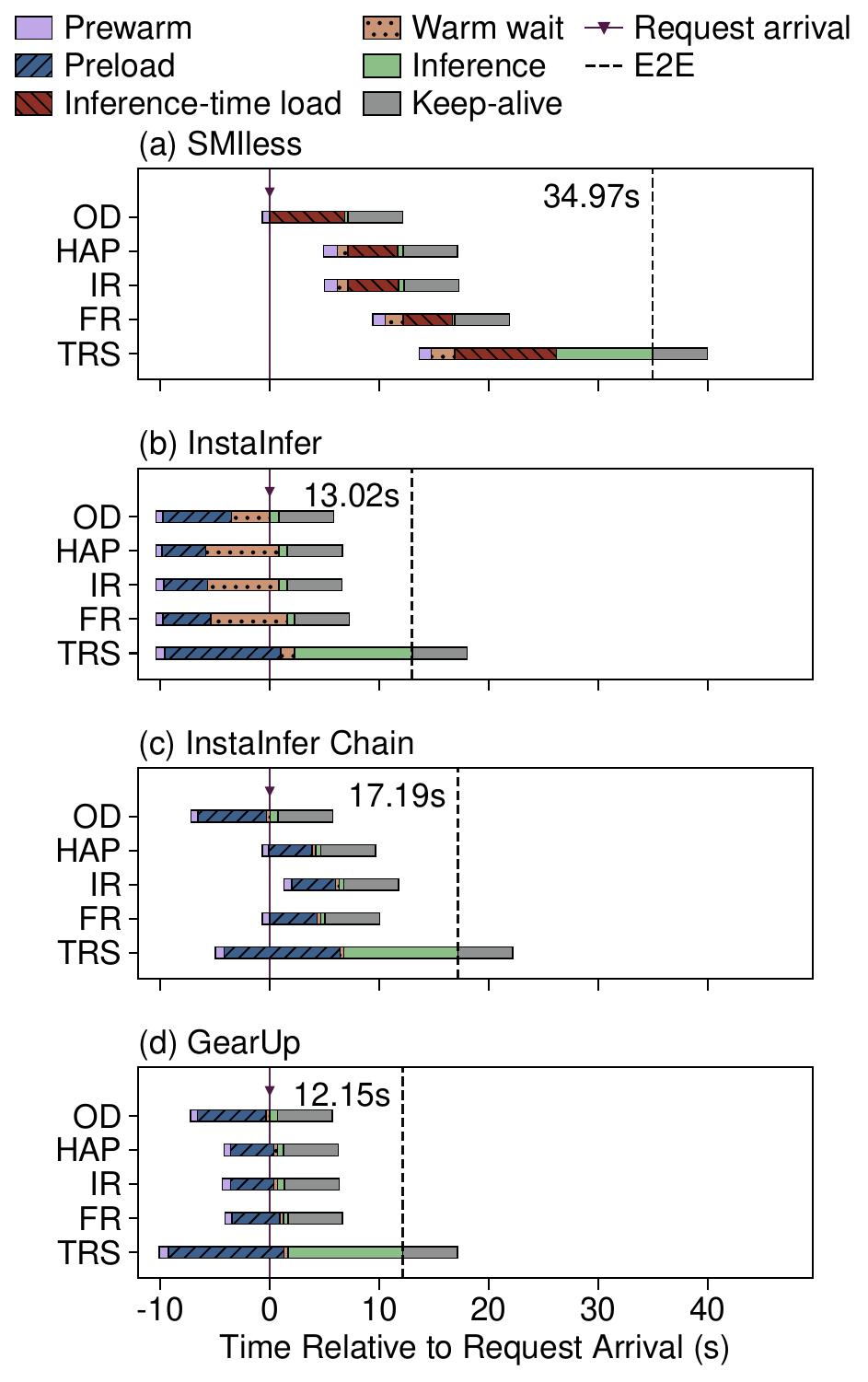}
    \vspace{-0.2in}
  \caption{Per-node execution timelines of AMBER Alert request \#317 under different serving approaches.}
    \vspace{-0.2in}

  \label{fig:mot-req317-timeline}
\end{figure}

\begin{figure}[t]
  \centering
  \includegraphics[width=0.7\linewidth]{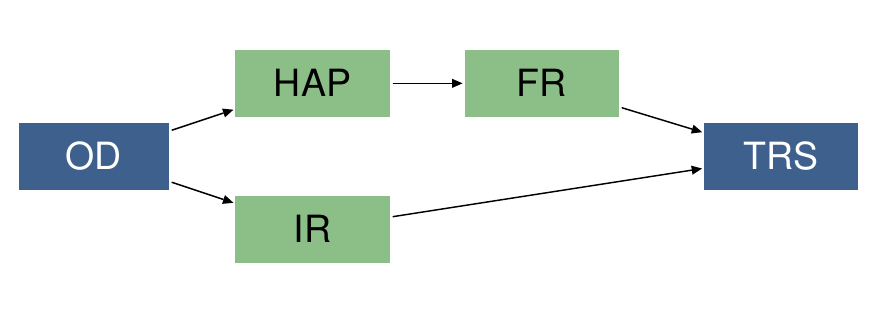}
  \vspace{-0.2in}
  \caption{
  DAG of AMBER Alert~\cite{romero2021llama,poms2018scanner}, a typical DAG-based ML inference workload for image-based emergency alerts. Each vertex denotes an ML function instance.
  }
    \vspace{-0.1in}

  \label{fig:amber-dag}
\end{figure}

\begin{figure}[t]
  \centering
  \includegraphics[width=\linewidth]{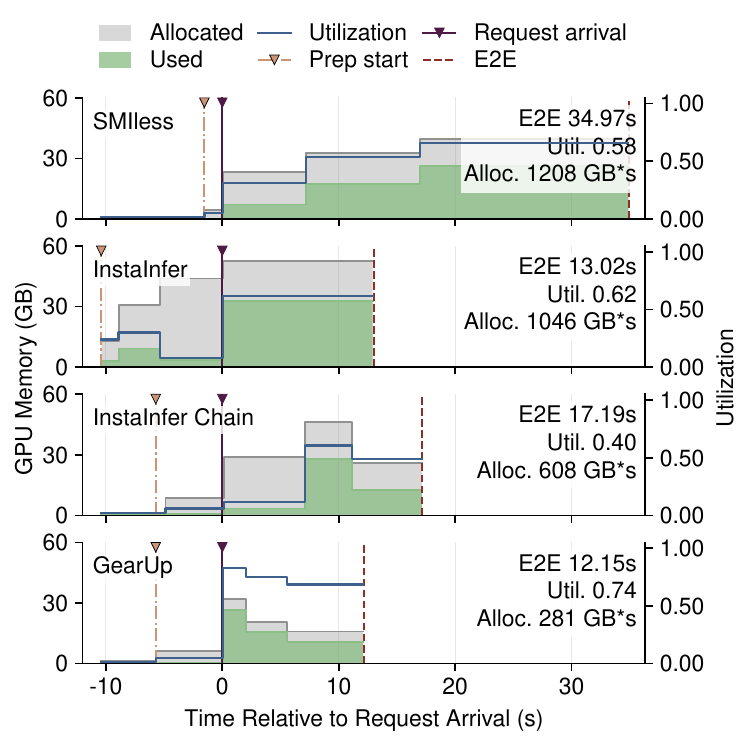}
    \vspace{-0.3in}
  \caption{GPU memory allocation and utilization for AMBER Alert request \#317.}
    \vspace{-0.2in}
  \label{fig:mot-req317-resource}
\end{figure}

The cost of ignoring this structure is large.
Figure~\ref{fig:mot-req317-timeline} traces one representative AMBER Alert request across four serving strategies on identical hardware, and Figure~\ref{fig:mot-req317-resource} reports the GPU memory each strategy holds.
SMIless~\cite{lu2024smiless}, a DAG-aware serving system, starts each function only after its dependencies finish, so ML loading stays on the request path and the request takes 34.97~s, most of it spent loading rather than performing inference.
InstaInfer~\cite{sui2024pre} instead pre-loads libraries and weights proactively; applied to the DAG as a single unit it reaches 13.02~s, but readies downstream nodes long before their dependencies finish and holds 1046~GB$\cdot$s of GPU memory largely idle.
InstaInfer Chain predicts each node's request arrival independently without DAG dependency information, causing late pre-loads and cascading stalls that leave GPUs idle while waiting for downstream initialization, giving 17.19~s at only 40\% GPU utilization.
\sys completes the same request in 12.15~s using 281~GB$\cdot$s.
Roughly two thirds of the reactive end-to-end latency is therefore initialization that a DAG-aware system can hide, and hiding it well costs a quarter of the memory eager pre-loading spends.

No existing system closes this gap.
Prior work does not jointly exploit DAG dependencies and fine-grained ML loading stages for pre-loading.
Single-function pre-loaders~\cite{rahimi2024pisel,sui2024pre} optimize ML loading without knowing which nodes a request will reach.
DAG serving systems~\cite{li2026iroute,bhasi2021kraken,mahgoub2022orion} schedule and size workflows without treating ML loading as a schedulable stage.
Table~\ref{tab:existingworks} and \S\ref{sec:background} compare these systems in detail.

Our key insight is that DAG dependencies expose opportunities to overlap downstream ML pre-loading with upstream execution, hiding initialization from the end-to-end critical path. However, realizing this opportunity raises three challenges.
\textbf{C1. Pre-loading must be timely and non-intrusive.}
ML nodes have different initialization costs.
Starting too late delays execution, while starting too early can interfere with active inference.
\textbf{C2. Pre-loading is costly within a request and redundant across requests.}
Eager preparation can leave substantial memory idle, while initializing the same models separately for each request repeats work.
\textbf{C3. Traffic and topology interact.}
Changing traffic and upstream progress make it difficult to decide which instances to retain and for how long. 

To bridge this gap, we propose \sys, a DAG-aware framework for serverless ML inference serving.
Only the critical path of a DAG determines its end-to-end latency, so every off-path node carries slack---time it spends waiting for its dependencies---and that slack is initialization time the request has already paid for.
\sys turns this observation into three mechanisms answering C1--C3, applied both within and across requests.

Our key contributions can be summarized as follows:
\begin{itemize}[topsep=0pt]
    \item \textbf{Slack-guided intra-request pre-loading (C1).} We traverse the DAG in reversed topological order to find the critical path and each node's slack $T_{slack}$, then schedule its loading stages inside that window, hiding initialization behind upstream execution without extending the critical path.
    \item \textbf{Execution-level instance reuse across requests (C2).} We decouple instance-level execution from the request-level DAG configuration, so warmed instances serve later requests that share the same model nodes.
    \item \textbf{DAG-aware traffic-adaptive keep-alive (C3).} A probabilistic cost model sets node lifetimes from DAG topology and observed arrivals, retaining reused nodes through bursts and propagating release decisions to dependents.
    \item \textbf{Implementation and evaluation.} We implement \sys across the controller, workers, function handlers, and DAG orchestrator, and evaluate it on the AMBER Alert and FINRA-ML workflows using NVIDIA L40S GPUs. \sys reduces median end-to-end latency by 64.5\% and 75.9\% against SMIless, cuts integrated GPU and host memory allocation by up to 83.9\% and 94.6\% against eager pre-loading, and achieves $4.2\times$ strong scaling on four workers.
\end{itemize}

\section{Background and Motivation}
\label{sec:background}
\subsection{Serverless ML Inference and Pre-loading}

A cold ML invocation must import libraries, load model weights, transfer them to GPU memory, and initialize CUDA kernels.
We call these four steps the node's \emph{loading stages}, and they are the unit \sys schedules.
For short inference tasks they dominate end-to-end latency, unlike generic serverless cold starts where container startup dominates.

Table~\ref{tab:existingworks} summarizes where existing systems fall.
One group is ML-aware but DAG-agnostic: InstaInfer~\cite{sui2024pre} pre-loads libraries and model artifacts for individual functions, and PISeL~\cite{rahimi2024pisel} pipelines download, deserialization, and execution within a function.
Both treat functions as independent units, so neither knows whether a pre-loaded node will be reached in the current request.
The other group is DAG-aware but loading-agnostic: Kraken~\cite{bhasi2021kraken}, Orion~\cite{mahgoub2022orion}, and iRoute~\cite{li2026iroute} optimize workflow scheduling, routing, and resource sizing, serverful systems such as Airflow~\cite{apache_airflow} and InferLine~\cite{crankshaw2020inferline} orchestrate DAG execution, and SMIless~\cite{lu2024smiless} co-optimizes resource configurations for ML DAGs but still launches each node reactively.
No system in either group treats the loading stages as work that can be scheduled ahead of invocation using DAG dependencies.
The consequence is a timing dilemma: loading too late exposes initialization on the execution path, while loading too early holds expensive GPU memory idle.

\subsection{Reactive DAG Execution Exposes ML Loading Overhead}
\label{sec:motivation-reactive}

DAG-aware scheduling alone does not hide ML loading when node initialization remains reactive. A downstream node waits for its dependencies before starting its own loading stages, causing initialization delays to accumulate along the request path.

Figure~\ref{fig:mot-req317-timeline} illustrates this dilemma on one representative AMBER Alert request, executed by all four approaches under the configuration of \S\ref{sec:setup}.
The DAG contains five GPU-backed functions: Object Detection (OD), Human Attribute Prediction (HAP), Image Recognition (IR), Face Recognition (FR), and Transcription (TRS), with the dependencies in Figure~\ref{fig:amber-dag}.

Figure~\ref{fig:mot-req317-timeline}(a) shows SMIless, which starts each function only after its dependencies become ready.
This avoids pre-loading and its resource management complexity, but leaves the loading stages on the request path.
The request spends most of its time loading rather than performing inference, yielding 34.97~s.
Hiding that cost requires starting downstream initialization before its predecessors finish.

\subsection{DAG-Unaware Pre-loading Can Be Wasteful or Late}
\label{sec:motivation-dag-unaware}



Pre-loading the entire DAG as one unit can prepare downstream instances too early.
In the Instainfer setup, we treat the whole request as one function and let InstaInfer pre-load every ML instance on arrival.
As a result, Figure~\ref{fig:mot-req317-timeline}(b) shows that this setup reduces latency to 13.02~s.
However, downstream instances become ready long before their dependencies finish, holding containers and GPU memory through long warm-wait periods (Figure~\ref{fig:mot-req317-resource}(b)).
When several requests compete for limited GPUs, this over-provisioning contends with the instances actually executing.

The second setup instead predicts each node's pre-load time independently.
In the Instainfer Chain setup (InstaInfer Chain), each node is treated as an independent function, with its pre-load time predicted without dependency information.
Figure~\ref{fig:mot-req317-timeline}(c) shows the opposite inefficiency: a late HAP pre-load delays FR, the delay cascades to TRS, and the late TRS pre-load extends the critical path to 17.19~s.

\begin{table}[t]
\centering
\small
\caption{Comparison of ML and DAG serving approaches. GearUp uniquely combines ML-aware pre-loading with DAG dependencies to proactively determine when downstream ML instances should be loaded.}
\label{tab:existingworks}
\begin{tabular}{@{}lcccc@{}}
\toprule
\textbf{Method} &
\shortstack{\textbf{DAG-}\\\textbf{Aware}} &
\shortstack{\textbf{ML-}\\\textbf{Aware}} &
\shortstack{\textbf{Proactive}\\\textbf{Pre-load}} &
\shortstack{\textbf{DAG-Guided}\\\textbf{Pre-load}} \\
\midrule
PISeL~\cite{rahimi2024pisel}
    & \xmark & \cmark & \xmark & \xmark \\
InstaInfer~\cite{sui2024pre}
    & \xmark & \cmark & \cmark & \xmark \\
iRoute~\cite{li2026iroute}
    & \cmark & \xmark & \xmark & \xmark \\
Kraken~\cite{bhasi2021kraken}
    & \cmark & \xmark & \xmark & \xmark \\
Orion~\cite{mahgoub2022orion}
    & \cmark & \xmark & \xmark & \xmark \\
SMIless~\cite{lu2024smiless}
    & \cmark & \cmark & \xmark & \xmark \\
\textbf{GearUp}
    & \cmark & \cmark & \cmark & \cmark \\
\midrule
\multicolumn{5}{@{}l}{\textit{Serverful DAG serving}} \\
Airflow~\cite{apache_airflow}
    & \cmark & \na & \na & \na \\
InferLine~\cite{crankshaw2020inferline}
    & \cmark & \cmark & \na & \na \\
\bottomrule
\end{tabular}
\end{table}

\subsection{Overlapping Pre-loading with Upstream Execution}
\label{sec:motivation-dag-awareness}

The above observations highlight the need to align pre-loading with DAG progress.
Our key insight is that DAG dependencies expose opportunities to overlap downstream ML pre-loading with upstream execution, hiding initialization from the end-to-end critical path.
Using these dependencies and execution slack to place loading stages allows downstream instances to be prepared before invocation without keeping them ready throughout upstream execution.

Figure~\ref{fig:mot-req317-timeline}(d) shows \sys, which uses DAG dependencies and execution slack to place pre-loading.
Downstream instances such as FR and TRS are prepared before they become executable, but without the warm-wait periods of Instainfer pre-loading, giving 12.15\,s.
Figure~\ref{fig:mot-req317-resource} quantifies the resource cost of reaching those latencies.
SMIless and Instainfer allocate 1208 and 1046~GB$\cdot$s of GPU memory, and Instainfer Chain reduces allocation to 608~GB$\cdot$s but reaches only 40\% utilization because its pre-loads are poorly timed.
\sys uses 281~GB$\cdot$s at 74\% utilization, the lowest allocation and the highest utilization of the four.

\subsection{Design Implications}
\label{sec:motivation-implications}

\begin{figure*}[t]
  \begin{center}
  \includegraphics[width=.98\textwidth]{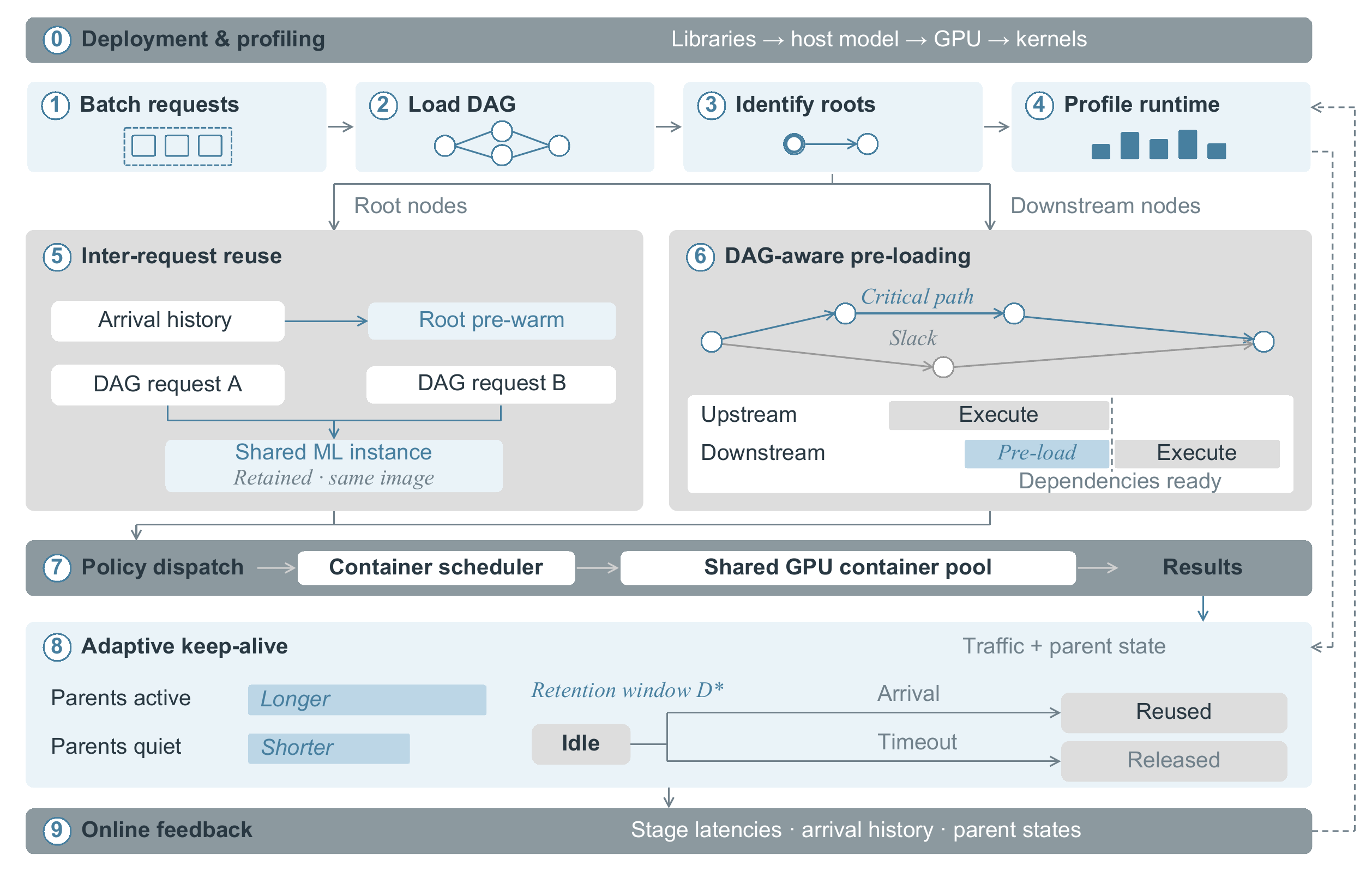}
  \vspace{-0.2in}
  \caption{Overview of \sys. Deployment-time profiling (0) precedes admission (1--3), per-node policy generation for root and downstream nodes (4--6), and execution with online feedback (7--9).}
  \label{fig-overview}
  \end{center}
\end{figure*}

Realizing this opportunity requires coordinating loading within a request and managing prepared instances across requests.
This raises the three challenges introduced in \S\ref{sec:intro}.

\textbf{C1: pre-loading must be timely and non-intrusive.} 
Unlike lightweight CPU function nodes, ML nodes have heterogeneous initialization costs, so some nodes must start loading several stages before execution.
The schedule must use DAG dependencies to find this window while avoiding interference with active inference.

\textbf{C2: pre-loading is costly within a request and redundant across requests.}
Preparing every reachable node can hold substantial resources idle or waste preparation on unused branches.
The same model nodes also recur across requests, making repeated initialization unnecessary.

\textbf{C3. Traffic and topology interact.}
DAG fan-out can amplify changes in root traffic into larger changes in downstream demand.
Static schedules and per-node policies can therefore retain too many instances or release them too early.

\section{An Overview of \sys}

\subsection{Objectives and Challenges}

\sys pursues two objectives: 1) minimize end-to-end inference latency while preserving the DAG's dependency constraints, and 2) avoid redundant initialization so that containers and GPU memory are not held for nodes that will not run soon.
Meeting them requires addressing challenges C1--C3 of \S\ref{sec:motivation-implications}: placing multi-stage pre-loading inside DAG slack, controlling its resource cost within and across requests, and coordinating node lifecycles under dependency-induced traffic.

\subsection{\sys's System Architecture}

\sys rests on one observation: only the critical path of a DAG determines its end-to-end latency, so every off-path node carries slack that the request is already spending.
Initialization placed inside a node's slack window is therefore free, while the same work placed outside it either delays the request or holds resources before they are needed.
Before any of this can be scheduled, \sys makes initialization schedulable: each container exposes its loading stages---library import, weight load into host memory, GPU transfer, and CUDA-kernel warm-up---as separately invocable, idempotent steps, and a profiler records the latency of each stage per node (\S\ref{subsec:pipeline-profiler}).
Over an inference DAG $G(V,E)$ annotated with these profiles, \sys then applies three components, one per challenge:

\begin{itemize}[topsep=0pt]
    \item \textbf{Intra-Request DAG-Aware Pre-Loader} (C1, \S\ref{sec:intra-request-preloader}) computes slack from a reversed topological traversal and overlaps each node's loading stages with upstream execution.
    \item \textbf{Inter-Request Batching and Reuse Scheduler} (C2, \S\ref{sec:inter-request-scheduler}) batches requests by DAG type and reuses unreleased instances across requests and DAGs that share model images.
    \item \textbf{Dynamic Keep-Alive Manager} (C3, \S\ref{sec:dynamic-keep-alive}) sets per-node retention from a topology-aware cost model over observed arrivals.
\end{itemize}

\subsection{\sys's Workflow}
\label{subsec:gearup-workflow}

Figure~\ref{fig-overview} shows how these components interact, in four phases; the numbers match the callouts in the figure.

\textbf{Deployment-time profiling (0).}
Developers deploy a DAG job with its code and model configuration.
Before serving, \sys runs pilot executions over sampled batch sizes to build the per-node stage-latency profiles used by the pre-loading pipeline of \S\ref{subsec:pipeline-profiler}.

\textbf{Admission (1--3).}
The front end groups arriving requests by DAG type within a time window, amortizing scheduling cost and exposing reuse opportunities across similar workloads (\S\ref{subsec:request-batching-reuse}).
For each batch it loads the DAG metadata and walks the nodes, separating root nodes from downstream nodes.
This split decides which policy applies: a root node has no upstream execution to hide behind, so its initialization must be predicted from arrival history, whereas a downstream node can be hidden inside DAG slack.

\textbf{Policy generation (4--6).}
The workload extractor and online profiler supply per-node execution times, stage latencies, and resource usage (\S\ref{subsec:pipeline-profiler}).
For a root node, a histogram over inter-arrival times decides when to pre-warm and whether a live instance can be reused (\S\ref{subsec:inter-request-histogram-prediction}, \S\ref{subsec:request-batching-reuse}).
For a downstream node, reversed-topology analysis yields the critical path and $T_{\text{slack}}$, and the profiled stage latencies determine where inside that window each stage is issued (\S\ref{subsec:critical-path-identification}, \S\ref{subsec:slack-guided-preloading}).

\textbf{Execution and feedback (7--9).}
Policies reach the workers as pre-load, request, reuse, and keep-alive commands, and the container scheduler enforces them over a shared GPU container pool, where an idle instance is either reused by an arriving request or released on timeout (\S\ref{subsec:container-lifecycle}).
The retention window $D^*$ is set per node by the cost model and conditioned on DAG parent activity, so active parents lengthen it and quiet parents shorten it (\S\ref{subsec:keep-alive-optimization}, \S\ref{subsec:idle-wait-estimation}, \S\ref{subsec:dag-aware-keepalive-adjustment}).
Runtime observations then update the stage latencies, arrival history, and parent states that these policies read, so they track changing traffic instead of staying fixed.

\section{Intra-Request DAG-Aware Pre-Loader}\label{sec:intra-request-preloader}

Existing systems either pre-load all nodes indiscriminately or rely on generic execution models that ignore ML-specific initialization pipelines.
\sys instead uses DAG topology and slack to defer each node's initialization to the latest point that still hides it.

\subsection{Per-Node Multi-Stage Pre-loading Pipeline and Profiler}\label{subsec:pipeline-profiler}

Each function container exposes a unified \texttt{/handle} endpoint that supports a staged, idempotent initialization pipeline:

\begin{enumerate}
    \item \texttt{preload\_libs}: import framework libraries (e.g., PyTorch).
    \item \texttt{preload\_model\_container}: load model weights into host memory.
    \item \texttt{preload\_model\_gpu}: transfer the model to GPU.
    \item \texttt{preload\_cuda\_kernels}: execute a dummy forward pass to trigger kernel initialization.
\end{enumerate}

Idempotence lets the scheduler issue the stages one at a time and stop after any of them without leaving the container in a partial state.
The scheduler orchestrates these stages sequentially via the endpoint, and stage latencies are recorded and persisted as JSON profiles for runtime scheduling.

\subsection{Critical Path Identification via Reversed Topology}\label{subsec:critical-path-identification}

\begin{figure}[t]
  \begin{center}
  \includegraphics[width=\linewidth]{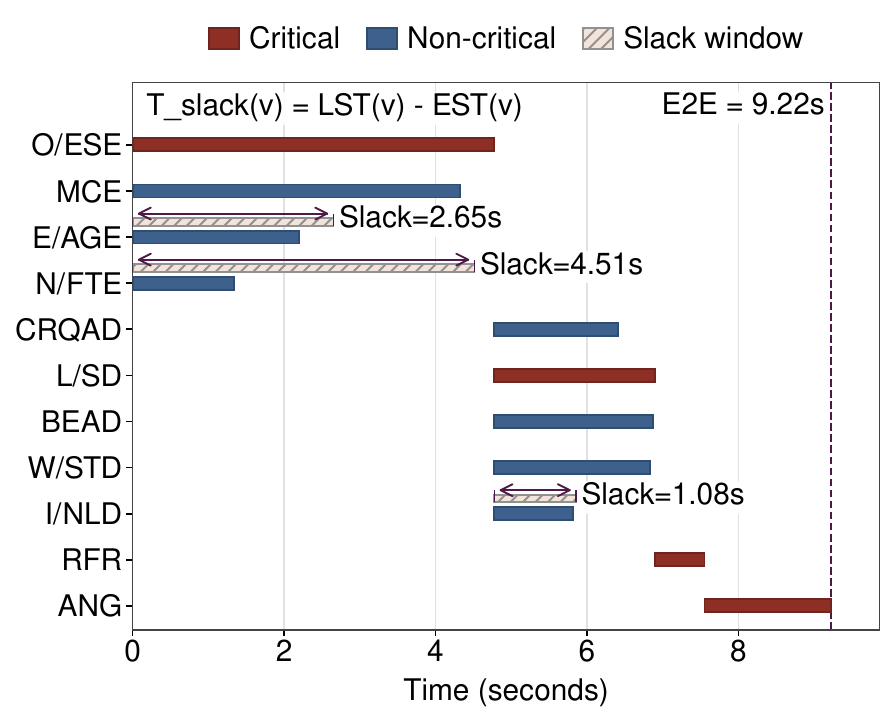}
  \vspace{-0.3in}
  \caption{EST/LST schedule and slack windows for FINRA-ML. Bars show per-node invocation durations.}
\label{fig:critical-path-slack-finra}
  \end{center}
\end{figure}

As shown in Figure~\ref{fig:critical-path-slack-finra}, upon receiving a request the scheduler uses profiled inference durations to compute earliest start times in topological order and latest start times in reversed topological order, identifying the inference-only critical path.
For each node $v_i$, we compute its slack time:
\[
    T_{slack}(v_i) = LST(v_i) - EST(v_i),
\]
where $EST(v_i)$ and $LST(v_i)$ denote the earliest and latest inference start times relative to request arrival under this profiled schedule.
Nodes on the critical path have $T_{slack}=0$, while non-critical nodes expose additional temporal flexibility for deferred initialization.
Because $EST(v_i)+T_{slack}(v_i)=LST(v_i)$, the latest start time is the deadline that preserves the profiled critical-path latency.
\sys targets this deadline when scheduling pre-loading, exploiting both upstream execution time and any additional slack.

\subsection{Slack-Guided Pre-loading}\label{subsec:slack-guided-preloading}

The orchestrator implements DAG-aware pre-loading from the identified critical path and the profiled latencies.
For each node $v_i$, $T_{\text{pre}}(v_i)$ is the sum of its four sequential pre-load stage latencies, and $T_{\text{inf}}(v_i)$ is its inference time.
Nodes are processed in topological order.
For node $v_i$, let $T_{\text{dep}}(v_i)$ denote the maximum estimated completion time of its parents, or zero for a root, and $t_i$ its pre-load start time relative to request arrival:
\[
t_i=\max\bigl(0,\;EST(v_i)+T_{slack}(v_i)-T_{\text{pre}}(v_i)\bigr),
\]
\[
T_{\text{end}}(v_i)=\max\bigl(T_{\text{dep}}(v_i),\;t_i+T_{\text{pre}}(v_i)\bigr)+T_{\text{inf}}(v_i).
\]
Pre-loading therefore finishes by $LST(v_i)$ whenever feasible; the clamp at zero marks nodes whose loading cannot be fully hidden, and for those the completion estimate absorbs the resulting delay.
Deferring initialization to this deadline, rather than issuing it at request arrival, is what separates \sys from eager pre-loading: a container is held only over the interval it is needed, which is what lowers integrated GPU memory allocation.
The cost is that a node keeps no margin of its own, so an underestimated $T_{\text{pre}}(v_i)$ extends the critical path; the online profiler refreshes these estimates from observed executions to keep that error small.
Pre-load and inference run in separate threads, synchronized by dependency completion and pre-load readiness, and a root node's pre-load stage is skipped when it is already warm.

\section{Inter-Request Batching and Reuse Scheduler}\label{sec:inter-request-scheduler}

\subsection{Inter-Request Preloading via Histogram Prediction}\label{subsec:inter-request-histogram-prediction}

For recurring DAGs, GearUp predicts inter-arrival times using histogram-based estimation. From historical traces, we derive an empirical distribution and compute a percentile-based estimate $\hat{t}$. The preload trigger for the root node, measured from the preceding root request's arrival, is:
\[
T_{\mathrm{trigger}}=\max\left(0,\widehat{t}-T_{\mathrm{pre}}(\mathrm{root})\right).
\]
At this time, the root is preloaded in anticipation of the next invocation.

\subsection{Request Batching and Instance Reuse Across Requests}\label{subsec:request-batching-reuse}

\textbf{Request Batching.} A batching time window is employed to reduce per-request overhead. Upon the first request for the same DAG, a batching window (default 500\,ms) is initiated, during which subsequent requests are aggregated. In parallel, the dispatcher prepares execution by either reusing an idle container or launching a new one. After the batching window expires, the aggregated requests are dispatched in a single call, which forwards them to the container for batched execution.

\textbf{Instance Reuse.} The framework decouples reuse policy from execution mechanism. Two nodes are interchangeable when they declare the same container image and model configuration, so an instance already initialized for one is in the state the other requires. Such nodes recur both across successive requests to the same DAG and across different DAGs that embed the same model, and in either case an idle instance is assigned instead of initializing a new one. The controller makes this decision because it maintains the pool across workers (\S\ref{subsec:container-lifecycle}); the scheduler's intra-DAG timing logic is unchanged, and a reused instance simply skips the pre-load stages of \S\ref{subsec:pipeline-profiler}. Reuse and retention are therefore coupled: an instance can only be reused while it is still being kept alive, so the retention window $D^*$ of \S\ref{sec:dynamic-keep-alive} determines how much reuse is available, and each reuse hit is a cold start and a model load the request does not pay for.

\section{Dynamic Keep-Alive Manager}\label{sec:dynamic-keep-alive}

\begin{figure}[t]
  \begin{center}
  \includegraphics[width=\linewidth]{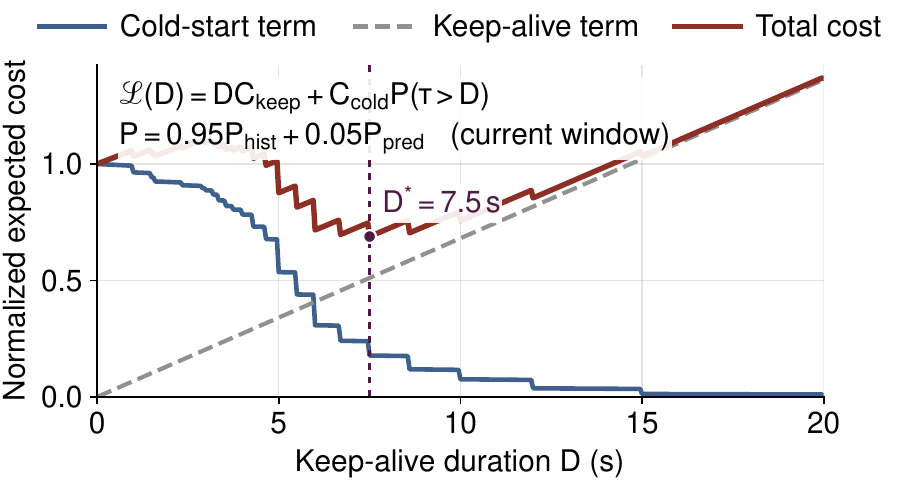}
  \vspace{-0.3in}
  \caption{Probabilistic keep-alive optimization for AMBER Alert with $w=0.95$.}
  \label{fig-dynamic}
  \end{center}
\end{figure}

Static timeout-based keep-alive policies are insufficient under bursty or irregular workloads and ignore DAG topology. As shown in Figure~\ref{fig-dynamic}, we instead formulate keep-alive as a probabilistic cost optimization problem.

\subsection{Container Lifecycle}\label{subsec:container-lifecycle}

Containers are launched with GPU support and mapped to dynamically assigned host ports. Workers maintain container metadata (function, port, status, timestamps) in a thread-safe structure. Upon request, idle containers are reused when available; otherwise, new containers are created. The controller coordinates reuse across workers and enforces global limits via eviction of the oldest idle containers.

\subsection{Cost-Benefit Keep-Alive Optimization}\label{subsec:keep-alive-optimization}

For a candidate keep-alive duration $D$, measured from when a container becomes idle, we use the following surrogate cost objective:
\[
    \mathcal{L}(D)
    = \underbrace{D \cdot C_{keep}}_{\text{Keep-Alive Cost}}
    + \underbrace{C_{cold} \cdot \mathbb{P}(\tau > D)}_{\text{Expected Cold-Start Cost}},
\]
where $C_{keep}$ is the unit cost per second of keeping an idle instance alive, $C_{cold}$ is the cold-start penalty expressed in the same cost units, and $\tau$ is the waiting time from the start of the idle period to the next invocation. The retention term charges the full timeout $D$, approximating the idle-memory cost. We estimate $\mathbb{P}(\tau > D)$ using a histogram and a short-term predictor.

The optimal keep-alive duration under this objective is:
\begin{equation}
    D^*
    = \arg\min_{D \geq 0}
    \left(
        D \cdot C_{keep}
        + C_{cold} \cdot \mathbb{P}(\tau > D)
    \right).
    \label{eq:keepalive-opt}
\end{equation}

Extending $D$ raises the retention term but lowers the expected cold-start term, since $\mathbb{P}(\tau > D)$ is non-increasing in $D$. $D^*$ is the point at which the two balance.

\subsection{Idle-Wait Probability Estimation}\label{subsec:idle-wait-estimation}

To estimate $\mathbb{P}(\tau > D)$, we combine historical observations and short-term prediction:
\[
    \mathbb{P}(\tau > D)
    = w \cdot P_{hist}(\tau > D)
    + (1-w) \cdot P_{pred}(\tau > D).
\]

The weight $w$ adapts dynamically:
\begin{itemize}
    \item $w \rightarrow 1$ under stable traffic,
    \item $w \rightarrow 0$ when sudden workload shifts are detected.
\end{itemize}

\textbf{Histogram-Based Component.}
$P_{hist}$ is derived from recent observations of the waiting time from idle onset to the next invocation. Instead of relying on a single average, we maintain bucketed frequency counts and compute:
\[
P_{hist}(\tau > D) = \frac{\sum_{\tau_j > D} \text{freq}(\tau_j)}{\sum_{\tau_j} \text{freq}(\tau_j)}.
\]
As an additional burst-handling heuristic, if dense arrivals are detected within a short interval, the system enforces a minimum non-zero $D$ to bridge burst gaps.

\textbf{Predictive Component.}
$P_{pred}$ reflects short-term forecasting. It can be modeled as a step function where the probability is 1 if the predicted waiting time exceeds $D$, and 0 otherwise.

This hybrid mechanism combines historical stability with short-term responsiveness.

\subsection{DAG-Aware Adjustment}\label{subsec:dag-aware-keepalive-adjustment}

Unlike DAG-unaware approaches, we condition keep-alive decisions on DAG structure. The waiting time until node $v_i$ receives its next invocation depends on the state of its parent nodes. 
We refine the probability as:
\[
    P'(\tau_i > D)
    = P(\tau_i > D \mid \mathcal{S}_{parent}),
\]
where $\mathcal{S}_{parent}$ represents the runtime state of predecessor nodes. For each node $v_i$, we substitute this conditional probability into Eq.~\eqref{eq:keepalive-opt} to compute its keep-alive duration $D_i^*$.

Parent activity updates the predicted distribution of the next invocation, and the cost model recomputes $D_i^*$ accordingly to balance retention cost against cold-start risk. 
Through this DAG-aware probabilistic control, the system adapts container retention to both intra-DAG execution signals and inter-request traffic patterns.
\section{Implementation and Evaluation}
\subsection{Implementation}

We implement \sys as a layer that can be integrated into serverless inference serving platforms, spanning the controller, workers, function handlers, and DAG orchestrator.
The system runs on GPU-enabled nodes, where each ML function is packaged as a Docker container exposing a unified HTTP interface.

\sys follows a centralized-controller, distributed-worker design.
The \textbf{controller} is the entry point, exposing a REST API through Flask, holding global container state in a thread-safe map, and enforcing a configurable upper bound on GPU memory.
DAG specifications are given as JSON adjacency lists, loaded at startup and broadcast to all workers.
Each \textbf{worker} manages container lifecycles on its node through the Docker SDK, probing readiness over HTTP with exponential backoff and reporting status to the controller.
A worker also runs the keep-alive optimizer of \S\ref{sec:dynamic-keep-alive}, with a background thread reclaiming containers past their duration $D^*$.

Each function is profiled offline over sampled batch sizes to measure its four pre-load stages and a representative inference latency; profiles, DAG structures, and system parameters are externalized as JSON and YAML.

\subsection{Experiment Setup}
\label{sec:setup}

\noindent\textbf{Testbed.}
Our single-worker testbed uses one AWS EC2 g6e.4xlarge instance with 16 vCPUs, 128~GiB of host memory, and one NVIDIA L40S GPU with 48~GB of GPU memory. The instance hosts the client, controller, worker, and ML function containers. The worker maintains a shared container pool and creates, reuses, or releases containers according to controller decisions.

The scalability experiments use two identical instances, placing one, two, or four workers across them. One instance also hosts the controller, which keeps a global view of worker resources and assigns DAG nodes while preserving dependencies.

\begin{figure}[t]
  \centering
  \includegraphics[width=0.7\linewidth]{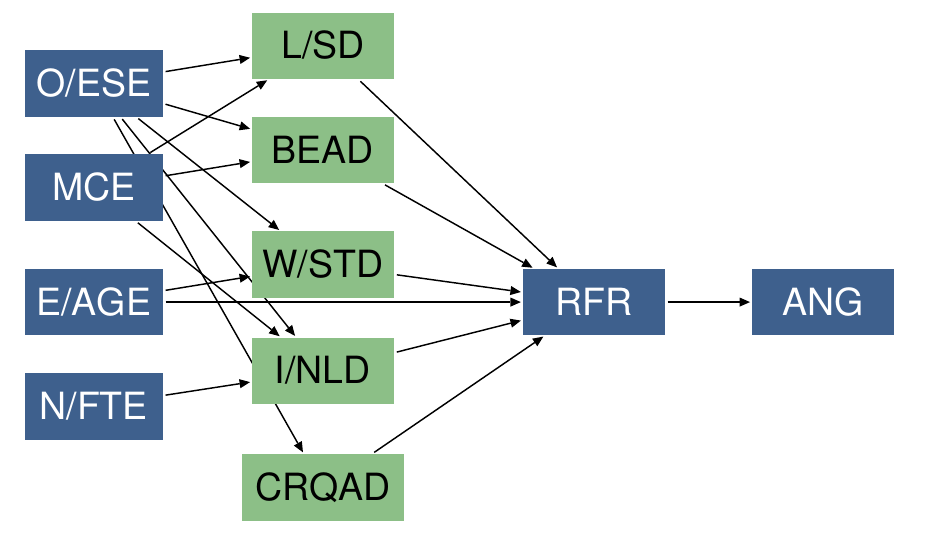}
  \vspace{-0.1in}
  \caption{
  DAG of FINRA~\cite{finra,li2026iroute}, a typical DAG-based ML inference workload for market surveillance. Each vertex denotes an ML function instance.
  }
  \label{fig:finra-dag}
\end{figure}
\begin{table}[t]
\centering
\footnotesize
\setlength{\tabcolsep}{3pt}
\caption{DAG nodes, functions, models, and datasets for the AMBER Alert and FINRA-ML workloads.}
\label{tab:dag-workloads}
\begin{tabular}{@{}l@{\hspace{4pt}}>{\raggedright\arraybackslash}p{0.355\linewidth}@{\hspace{4pt}}>{\raggedright\arraybackslash}p{0.405\linewidth}@{}}
\toprule
\textbf{Abbr.} & \textbf{Function} & \textbf{Model / Dataset} \\
\midrule
\multicolumn{3}{@{}l}{\textit{AMBER Alert}} \\
OD    & Object Detection            & YOLOv5 / COCO \\
HAP   & Human Attribute Prediction  & ResNet50 / ImageNet \\
FR    & Face Recognition            & FaceNet / ImageNet \\
IR    & Image Recognition           & ResNet50 / ImageNet \\
TRS   & Transcription               & T5 / SQuAD \\
\addlinespace[2pt]
\multicolumn{3}{@{}l}{\textit{FINRA-ML}} \\
O/ESE & Order/Event Sequence Encoder   & Transformer Encoder / CAT Order Events \\
MCE   & Market-Context Encoder         & Temporal Fusion Transformer / SIP and TAQ Market Data \\
E/AGE & Entity/Account Graph Encoder   & R-GCN / CAT Entity-Account Graph \\
N/FTE & News/Filing Text Encoder       & FinBERT / SEC EDGAR and Financial News \\
L/SD  & Layering/Spoofing Detector     & XGBoost / Labeled CAT Surveillance Cases \\
BEAD  & Best-Execution Anomaly Detector & LightGBM / CAT and NBBO Execution Records \\
W/STD & Wash/Self-Trade Detector       & GraphSAGE / CAT Trade-Account Graph \\
I/NLD & Inside/News Leakage Detector   & FinBERT + Transformer / CAT Trades and News Events \\
CRQAD & CAT Reporting-Quality Anomaly Detector & Isolation Forest / CAT Reporting Audit Logs \\
RFR   & Risk-Fusion Ranker             & LambdaMART / Historical Surveillance Alerts \\
ANG   & Alert-Narrative Generator      & T5 / Historical Alert Narratives \\
\bottomrule
\end{tabular}
\end{table}

\noindent\textbf{Workloads.} We evaluate \sys on the AMBER Alert and FINRA-ML workflows. AMBER Alert is the five-function DAG of Figure~\ref{fig:amber-dag}, where OD, HAP, IR, and FR use a batch size of 15 and TRS uses 512.

FINRA-ML is a market-surveillance workflow of eleven functions: four encoders feed five detectors, whose outputs a risk-fusion ranker aggregates before an alert-narrative generator. Figure~\ref{fig:finra-dag} gives its structure and Table~\ref{tab:dag-workloads} the models and datasets for both workflows.

\noindent\textbf{Request trace.}
We sample a continuous 300-minute segment from the Azure Functions Trace 2019 dataset~\cite{shahrad2020serverless}. The segment contains 319 root requests, with an average rate of 1.06 requests per minute. The ablation and scalability experiments each use a 160-request subset of this trace. In Figure~\ref{fig:traffic-adaptation}, we aggregate requests into 120-second windows. The maximum window contains eight root requests. We define a burst as a non-empty window containing at least three root requests, which corresponds to the 90th percentile of non-empty windows. The median inter-arrival time is 54.5~s, and the coefficient of variation is 0.516. The trace contains both busy and idle intervals, allowing us to evaluate instance reuse and adaptive keep-alive decisions under changing request pressure.

\noindent\textbf{Baseline systems.}
We compare \sys with three baselines. SMIless follows DAG dependencies but initializes downstream functions reactively and does not preload their ML components. InstaInfer eagerly prewarms and preloads all reachable ML functions without considering DAG execution timing. InstaInfer Chain applies function-local pre-loading along the invocation chain, with each function predicting its preload time independently.

The cold-execution reference $L_{\mathrm{cold}}$ defined above is 38.7~s for AMBER Alert and 92.6~s for FINRA-ML.

\begin{figure}[t]
    \centering
    \includegraphics[width=0.49\textwidth]
{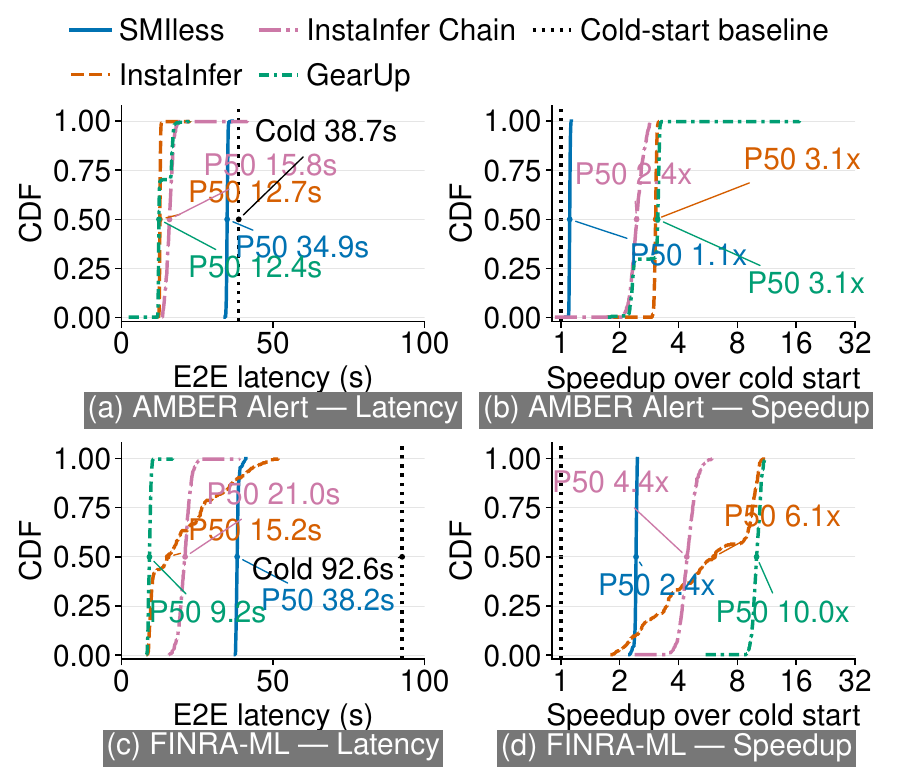}
\vspace{-0.3in}
  \caption{End-to-end latency and speedup distributions for AMBER Alert and FINRA-ML. Dotted lines mark the cold-execution references (latency and $1\times$ speedup).}
\label{fig:e2e-speedup}
\end{figure}

\subsection{Evaluation Metrics}

We use end-to-end latency, speedup over cold execution, resource utilization, and integrated resource allocation as the main evaluation metrics.

\noindent\textbf{End-to-end latency.}
End-to-end latency is measured from the submission of a DAG request until its sink node returns the final result. It includes container startup, library and model loading, GPU preparation, inference, communication, and dependency-induced waiting.
\begin{figure*}[t]
  \centering
  \includegraphics[width=\textwidth]{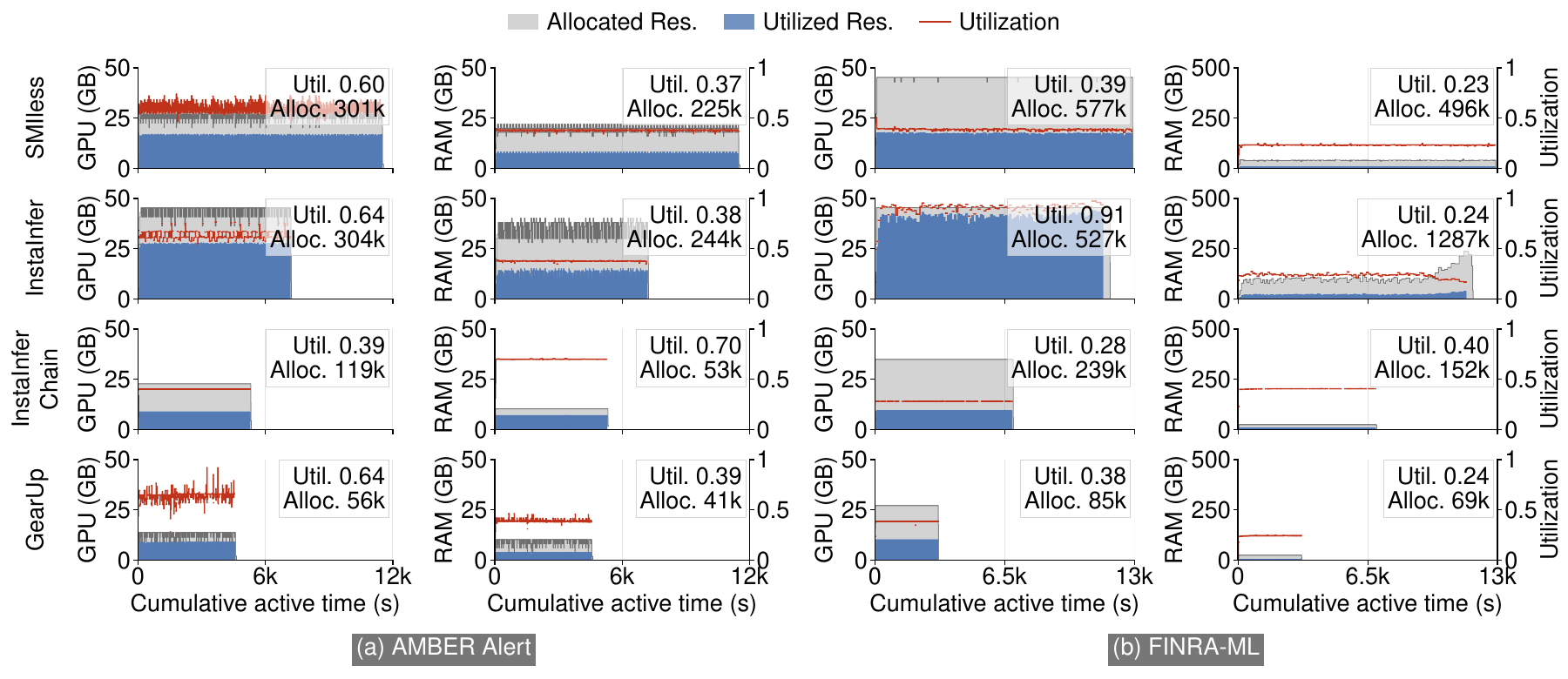}
  \vspace{-0.3in}
  \caption{GPU and host memory usage for (a) AMBER Alert and (b) FINRA-ML, excluding inter-request idle intervals.}
\label{fig:resource-utilization}
\vspace{-0.1in}
\end{figure*}


\noindent\textbf{Speedup over cold execution.}
For request $i$, speedup is defined as $\mathrm{Speedup}_i =
    \frac{L_{\mathrm{cold}}}{L_i}$, where $L_{\mathrm{cold}}$ is the cold-execution latency of the corresponding workload and $L_i$ is the end-to-end latency of request $i$. $L_{\mathrm{cold}}$ is a common unoptimized reference shared by all methods within a workload, obtained from a cold DAG traversal in which every node incurs container prewarming, model loading, and inference with no pre-loading, reuse, or keep-alive. Dependencies and parallel branches are preserved, so $L_{\mathrm{cold}}$ remains set by the critical path. A value greater than one indicates an improvement over cold execution.

\noindent\textbf{Resource utilization.}
For resource $r$, request-window utilization is defined as
\[
    U_r =
    \frac{
        \int_{\mathcal{W}} R_{\mathrm{used},r}(t)\,dt
    }{
        \int_{\mathcal{W}} R_{\mathrm{alloc},r}(t)\,dt
    },
\]
where $\mathcal{W}$ denotes request execution windows, $R_{\mathrm{used},r}(t)$ is the utilized resource, and $R_{\mathrm{alloc},r}(t)$ is the allocated resource. We report GPU memory and host memory utilization.

We also report integrated allocation,
\[
    A_r = \int R_{\mathrm{alloc},r}(t)\,dt,
\]
in $\mathrm{GB\,s}$. This metric captures both the amount of allocated memory and how long the allocation is retained. Memory plots summarize average utilization and integrated allocation, shown as \texttt{Util.}/\texttt{Alloc.} annotations or utilization/allocation pairs in row headings.


\subsection{Effectiveness of \sys}

\noindent\textbf{End-to-end latency and speedup.}
Figure~\ref{fig:e2e-speedup} compares the latency and speedup distributions of the four systems. For AMBER Alert, SMIless has a median latency of 34.9~s because downstream functions still incur initialization overhead. Eager pre-loading reduces the median to 12.7~s with InstaInfer, while function-local pre-loading achieves 15.8~s with InstaInfer Chain. \sys achieves the lowest median latency of 12.4~s, a 64.5\% reduction relative to SMIless. Based on the 38.7~s cold-execution baseline, the corresponding median speedups are $1.1\times$, $3.1\times$, $2.4\times$, and $3.1\times$, respectively.

The benefit is larger for FINRA-ML, whose deeper and wider DAG provides more opportunities for coordinated pre-loading and reuse. SMIless, InstaInfer, and InstaInfer Chain have median latencies of 38.2~s, 15.2~s, and 21.0~s, respectively. \sys reduces the median to 9.2~s, which is 75.9\% lower than SMIless and 39.5\% lower than InstaInfer. Based on the 92.6~s cold-execution baseline, its median speedup reaches $10.0\times$, compared with $2.4\times$ for SMIless, $6.1\times$ for InstaInfer, and $4.4\times$ for InstaInfer Chain.

\begin{figure}[t]
  \centering
  \includegraphics[width=\linewidth]{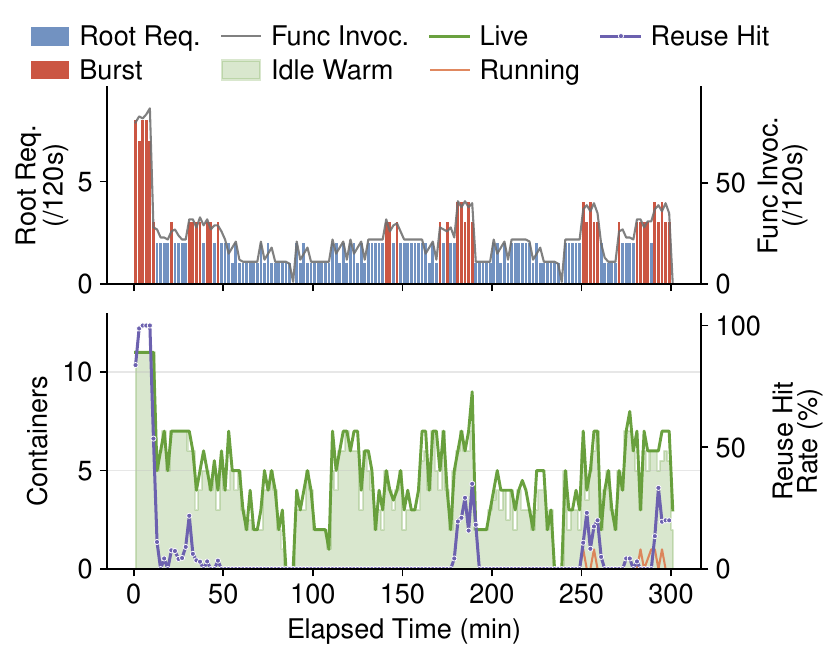}
  \vspace{-0.3in}
  \caption{Traffic-adaptive keep-alive behavior for FINRA-ML: request and invocation counts (top), and container counts and instance-reuse hit rate (bottom).}
  \vspace{-0.1in}
\label{fig:traffic-adaptation}
\end{figure}
\begin{figure}[t]
  \centering
  \includegraphics[width=\linewidth]
  {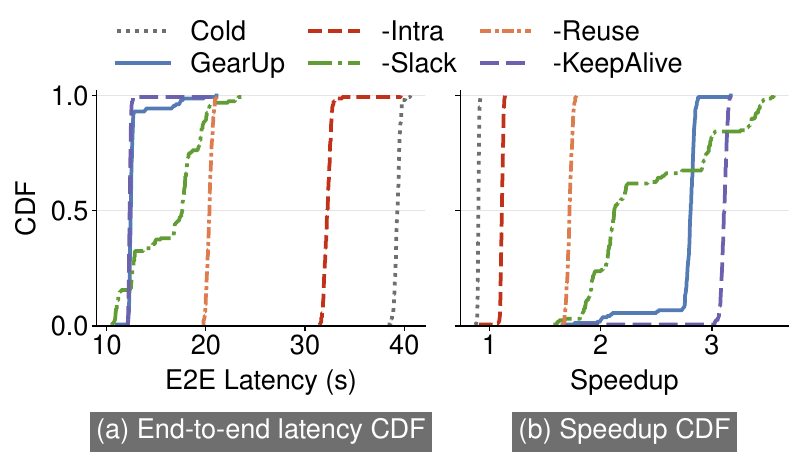}
  \vspace{-0.3in}
  \caption{AMBER Alert ablation results: (a) end-to-end latency CDF and (b) speedup CDF.}
\label{fig:ablation-performance}
\end{figure}

\begin{figure}[t]
  \centering
  \includegraphics[width=\linewidth]{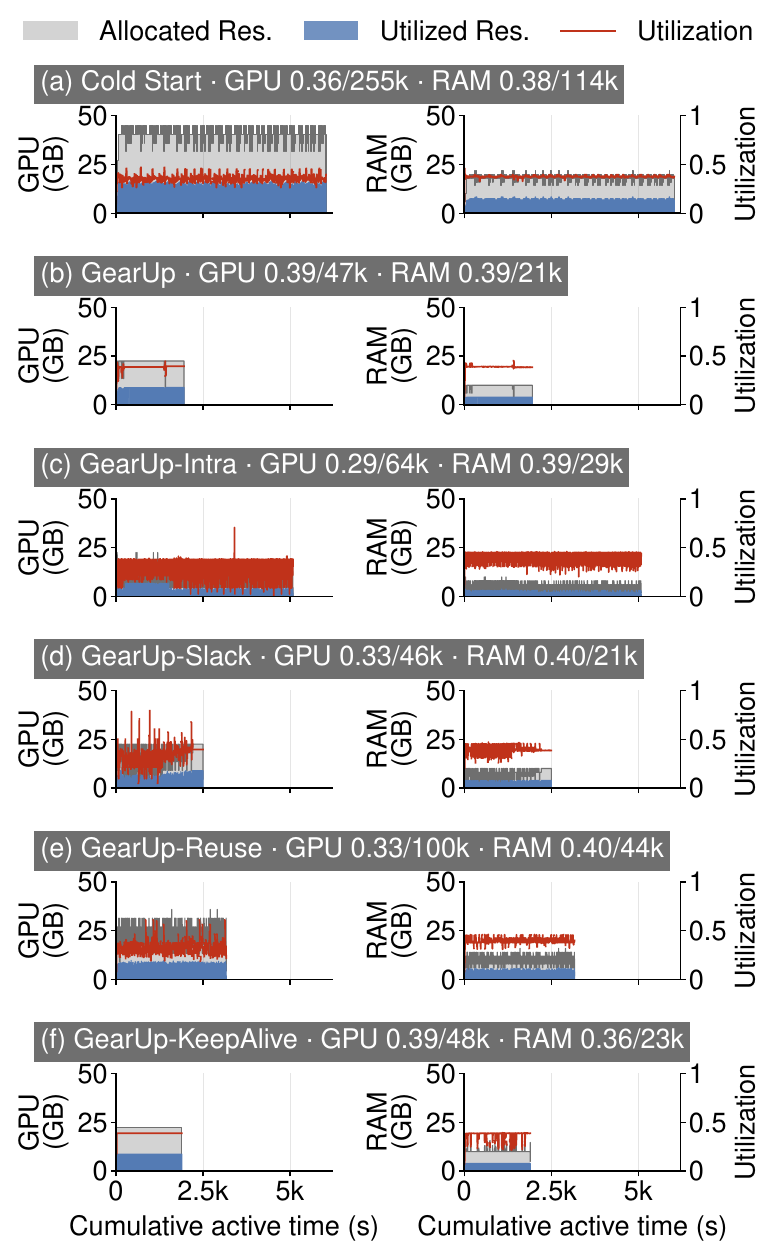}
  \vspace{-0.3in}
 \caption{GPU and host memory usage in the AMBER Alert ablation study, excluding inter-request idle intervals.}
\label{fig:ablation-resource}
\end{figure}

\noindent\textbf{Memory efficiency.} Figure~\ref{fig:resource-utilization} shows GPU and host memory allocation and utilization. For AMBER Alert, \sys cuts integrated GPU memory allocation from InstaInfer's 304k to 56k~$\mathrm{GB\,s}$ (81.6\%) and host memory from 244k to 41k~$\mathrm{GB\,s}$ (83.2\%). \sys retains the same average GPU utilization as InstaInfer, 0.64, while keeping resources allocated for a much shorter period.

For FINRA-ML, the reductions are 527k to 85k~$\mathrm{GB\,s}$ (83.9\%) and 1{,}287k to 69k~$\mathrm{GB\,s}$ (94.6\%). These results show that \sys reduces latency without keeping all DAG stages initialized throughout the workload.

\noindent\textbf{Traffic adaptation and instance reuse.} 
Figure~\ref{fig:traffic-adaptation} shows the keep-alive behavior of \sys under the FINRA-ML workload over the 300-minute trace. Although the number of root requests varies within a relatively narrow range outside the initial burst, each request fans out across the eleven-node DAG. This amplifies changes in root arrivals into larger variations in function invocation pressure.

The number of live containers changes with the resulting invocation demand, while a small idle-warm pool is retained across nearby windows. Reuse hits occur mainly during higher-load intervals, showing that recently initialized instances are reused when subsequent invocations arrive. During lower-load intervals, the live-container population decreases and few containers remain idle-warm. These results show that \sys adjusts container retention according to the invocation pressure produced by both external arrivals and DAG fan-out.

\subsection{Ablation Study}

We evaluate four \sys variants on the AMBER Alert workload. These runs use the 160-request subset of the trace, so their absolute allocations are lower than the full-trace figures reported above and should be compared only with each other. \textit{-Intra} disables intra-request pre-loading, \textit{-Slack} disables DAG slack-guided scheduling, \textit{-Reuse} disables instance reuse across requests, and \textit{-KeepAlive} disables adaptive keep-alive decisions.

\noindent\textbf{Latency impact.}
Figure~\ref{fig:ablation-performance} shows the latency and speedup distributions. Removing intra-request pre-loading causes the largest degradation, shifting latency close to cold execution and confirming that overlapping downstream initialization with upstream execution provides the main latency benefit. Removing instance reuse also increases latency because recurring requests must initialize new instances. Removing slack-guided scheduling produces a wider latency distribution, showing that DAG timing is needed to keep pre-loading work away from the active execution path. Disabling adaptive keep-alive has a smaller latency effect for this trace because many reusable instances are selected directly through inter-request reuse.

\noindent\textbf{Resource impact.}
Figure~\ref{fig:ablation-resource} reports GPU and host memory usage. Full \sys uses 47k~$\mathrm{GB\,s}$ of GPU memory allocation and 21k~$\mathrm{GB\,s}$ of host memory allocation. Without instance reuse, these values increase to 100k and 44k~$\mathrm{GB\,s}$, respectively, making reuse the largest contributor to resource savings. Disabling intra-request pre-loading increases GPU and host memory allocation to 64k and 29k~$\mathrm{GB\,s}$ and reduces average GPU utilization from 0.39 to 0.29.

Removing slack-guided scheduling results in 46k~$\mathrm{GB\,s}$ of GPU allocation and 21k~$\mathrm{GB\,s}$ of host memory allocation. Its main effect is therefore on latency rather than total allocation. Disabling adaptive keep-alive increases the allocations only slightly, to 48k and 23k~$\mathrm{GB\,s}$, under this workload. Together, the results show that intra-request pre-loading primarily reduces latency, while instance reuse provides the largest reduction in resource allocation.

\begin{figure}[t]
  \centering
  \includegraphics[width=\linewidth]{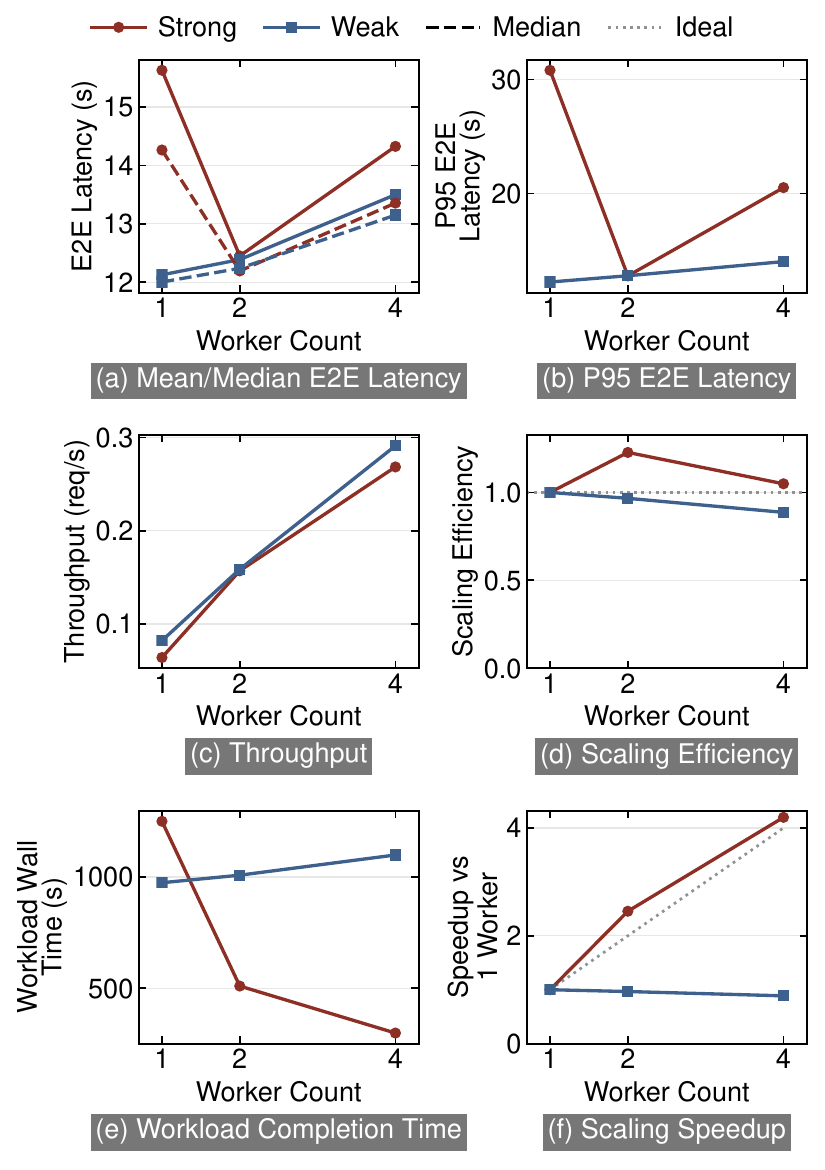}
  \vspace{-0.3in}
  \caption{Strong and weak scaling with one, two, and four workers. Dotted lines indicate ideal scaling.}
\label{fig:scalability}
\end{figure}

\subsection{Scalability}

We evaluate strong and weak scaling with one, two, and four workers. Strong scaling keeps the total workload fixed as workers are added. Weak scaling increases the workload in proportion to the worker count.

\noindent\textbf{Strong scaling.}
As shown in Figure~\ref{fig:scalability}, going from one to four workers raises throughput from 0.07 to 0.27 requests per second and cuts completion time from about 1{,}250~s to 300~s, a $4.2\times$ speedup at a scaling efficiency slightly above one. Two workers give the lowest per-request latency, with mean latency falling from 15.7~s to 12.4~s and P95 from 30.7~s to 12.8~s; at four workers both rise to 14.3~s and 20.5~s as coordination increases, still below the one-worker results.

\noindent\textbf{Weak scaling.}
Under weak scaling, throughput rises from 0.08 to 0.29 requests per second while mean latency stays between 12.1~s and 13.5~s and P95 grows only from 12.2~s to 14.0~s. Completion time grows from about 970~s to 1{,}100~s for a four-fold workload, a scaling efficiency near 0.9, so \sys holds per-request performance steady as the workload grows.
\section{Related Work}

\noindent\textbf{Serverful pipeline serving.}
InferLine~\cite{crankshaw2020inferline} profiles individual models and jointly tunes provisioning across pipeline stages, while Clipper~\cite{crankshaw2017clipper}, TFX~\cite{baylor2017tfx}, Airflow~\cite{apache_airflow}, Kubeflow~\cite{kubeflow}, and MLRun~\cite{mlrun} orchestrate ML DAGs in long-running clusters.
Because their services stay active, cold-start cost is largely irrelevant and fine-grained initialization control is unnecessary.

\noindent\textbf{Serverless DAG execution.}
Kraken~\cite{bhasi2021kraken}, Xanadu~\cite{daw2020xanadu}, DayDream~\cite{roy2022daydream}, Owl~\cite{tian2022owl}, and Orion~\cite{mahgoub2022orion} schedule and provision serverless workflows, but treat functions as generic tasks.
They do not model the large model artifacts, GPU residency, and multi-stage initialization that dominate ML inference startup.

\noindent\textbf{Cold-start mitigation.}
IceBreaker~\cite{roy2022icebreaker} keeps containers warm from workload prediction and Pagurus~\cite{li2021pagurus} shares containers across functions, but both treat a function as a black box with opaque startup cost.
InstaInfer~\cite{sui2024pre} and PISeL~\cite{rahimi2024pisel} target ML initialization directly, pre-loading libraries and weights or pipelining loading with inference; both assume a flat execution structure and cannot coordinate decisions across dependent stages.
SMIless~\cite{lu2024smiless} serves DAG-based ML inference but schedules and sizes reactively, without topology-aware pre-loading.

\noindent\textbf{Serverless LLM and agentic serving.}
ServerlessLLM~\cite{serverlessllm}, HydraServe~\cite{hydraserve}, and ServerlessLoRA~\cite{serverlesslora} reduce cold-start cost for LLM endpoints through checkpoint loading, model partitioning, and backbone sharing.
These optimize an individual model's lifecycle, whereas \sys targets the workflow layer above it.
Agent frameworks compose model calls, retrievers, and tools into graph-structured executions~\cite{langgraph,openai_agents_sdk} whose latency-critical segments are bounded ML DAGs, and \sys coordinates pre-loading and retention across those dependent functions rather than replacing model-level optimizations.

\section{Conclusion}
\label{sec:conclusion}

We presented \sys, a DAG-aware serverless inference system that coordinates initialization, execution, reuse, and retention across dependent ML stages.
\sys derives each node's slack from the DAG's critical path and defers its loading stages to the latest point that still hides them, reuses initialized instances across requests that share a model, and sets container lifetimes from a cost model conditioned on parent activity.
On the AMBER Alert and FINRA-ML workflows, \sys reduces median end-to-end latency by 64.5\% and 75.9\% against SMIless and cuts integrated GPU and host memory allocation by up to 83.9\% and 94.6\% against eager pre-loading, while sustaining a $4.2\times$ strong-scaling speedup on four workers.


\bibliographystyle{ACM-Reference-Format}
\bibliography{reference}

\end{document}